\documentclass[conference,10pt]{IEEEtran}
\usepackage{xcolor}
\usepackage{xcolor}
\usepackage{mathtools}
\usepackage{epsfig,color,subfigure}
\usepackage{amsmath,amssymb,amssymb,bbm,enumitem}
\usepackage{graphicx,lipsum}
\usepackage[switch,pagewise]{lineno}
\usepackage{hyperref}
\hypersetup{
        colorlinks = true,
        citecolor=red,
}
\def\cX{{\mathcal X}}

\def\uE{{\mathbb E}}

\def\be{ \begin{equation} }
\def\ee{ \end{equation} }
\def\bea{ \begin{eqnarray} }
\def\eea{ \end{eqnarray} }

\def\bx{{\bf x}}

\def\bz{{\bf z}}

\def\b0{{\bf 0}}

\def\cC{{\mathcal C}}

\def\cZ{{\mathcal Z}}
\def\cR{{\mathcal R}}

\def\cS{{\mathcal S}}
\def\cT{{\mathcal T}}

\def\sH{{\sf H}}
\def\sI{{\sf I}}

\newcommand*{\affaddr}[1]{#1} % No op here. Customize it for different styles.
\usepackage{subcaption}
\newcommand*{\affmark}[1][*]{\textsuperscript{#1}}

\begin{document}

\title{Semantic Communication Challenges: Understanding Dos and Avoiding Don'ts}
\author{%
Jinho Choi\affmark[1], Jihong Park\affmark[1], Eleonora Grassucci\affmark[2], and Danilo Comminiello\affmark[2] \\
\affaddr{\affmark[1]School of Information Technology, Deakin University,  Australia }\\
\affaddr{\affmark[2]Department of Information Engineering, Electronics and Telecommunication, Sapienza University of Rome, Italy}\\

}

\maketitle

\begin{abstract}
Semantic communication, emerging as a promising paradigm for data transmission, offers an innovative departure from the constraints of Shannon theory, heralding significant advancements in future communication technologies. Despite the proliferation of proposed approaches, there are still numerous challenges. In this paper, we review current semantic communication methodologies and shed light on pivotal issues and addressing certain discrepancies that exist within the field. By elucidating both dos and don'ts, we aim to provide valuable insights into the emerging landscape of semantic communication.
\end{abstract}

%\keywords{Semantic Communication, Wireless Caching}

\section{Introduction}
% what is semantic communication
% which are the advantages
% issues: challenges and misunderstandings (dos, donts)
% Our contribution

In the last years, communication literature has focused on the first level of Shannon and Weaver theory of communication, that is the technical level, mainly dealing with the technical aspects of the transmission. On the contrary, in recent years, a novel communication paradigm has emerged moving from the first to the second level of communication levels, namely semantic communication. The latter is based on the transmission and reconstruction of the transmitted content semantics (i.e., the meaning), without necessarily recovery the whole bitstream. By conveying semantic information only, semantic communication promises to reduce the required bandwidth and improve the efficiency of communication frameworks. Simultaneously, it completely revolutionizes the communication paradigm, opening the path to novel applications enabled by the marriage between semantic communication and generative AI. Generative models such as large language or diffusion models excel in regenerating content from semantic information, revealing to be groundbreaking tools for semantic communication \cite{Nam2023LanguageOrientedCW, Grassucci2023GenerativeSC, Grassucci2023DiffusionMF, Wijesinghe2023DiffGODG, Wu2023CDDMCD}.

Concurrently, as semantic communication interest is growing, novel challenges and issues to investigate are emerging, as well as some misunderstandings and avoidable practices. Sure enough, the direct comparison between semantic communication and conventional communication systems should be avoided, as they have different objectives and the restored content cannot be directly compared. In addition, we should refrain from thinking that semantic communication frameworks will replace conventional systems. Similarly, semantic communication was not conceived to be the paradigm that will be involved in every communication type within each system and infrastructure. Indeed, semantic communication should be rather envisaged as a novel communication model specifically tailored for some tasks or applications. In addition, especially in combination with generative models, semantic communication should be seen as a pathway to broaden the possibilities of communication systems for applications that were not even conceivable with conventional methods \cite{grassucci2024generative}. The applications that generative semantic communication may open are yet to be investigated.
Going further, a challenge this research topic unlocks is the investigation of security aspects in semantic communications, which is still widely unexplored. 

The scope of this paper is twofold. First, this paper aims to limit the misunderstandings that often come up around the discussions on semantic communication, specifically presenting the practices to avoid when working on this novel topic in terms of \textit{don'ts}. We believe that clarifying these aspects and delineating the dividing line between conventional systems and semantic ones may be crucial for the proliferation and development of future semantic communication research. Second, this paper also delineates the challenges and the issues to be further investigated in semantic communication in terms of \textit{dos}, including the role of generative AI and the security of such a paradigm. As semantic communication is an emerging topic, several pathways should be covered to guarantee safety and expand its fields of application.

The paper is organized as follows: Section~\ref{sec:semcom} presents the key background on semantic communication, in Section~\ref{sec:donts} we discuss the \textit{don'ts}, that are the practices to avoid, while challenges and opportunities (the \textit{dos}) are debated in Section~\ref{sec:dos}. Finally, conclusions are drawn in Section~\ref{sec:conc}.

% \cite{Habibian19}

% do
% \begin{itemize}
%     \item Exploit Generative AI ability to regenerate content from the semantic information 
% \end{itemize}

% don'ts
% \begin{itemize}
%     \item Direct comparisons between TC and SC are invalid as they have different purposes.
%     \item Think that semantic communication should replace every other communication system
% \end{itemize}

\section{Semantic Communication}
\label{sec:semcom}

\subsection{Existing Approaches}

\begin{figure}[h]
\begin{center}
\includegraphics[width=\columnwidth]{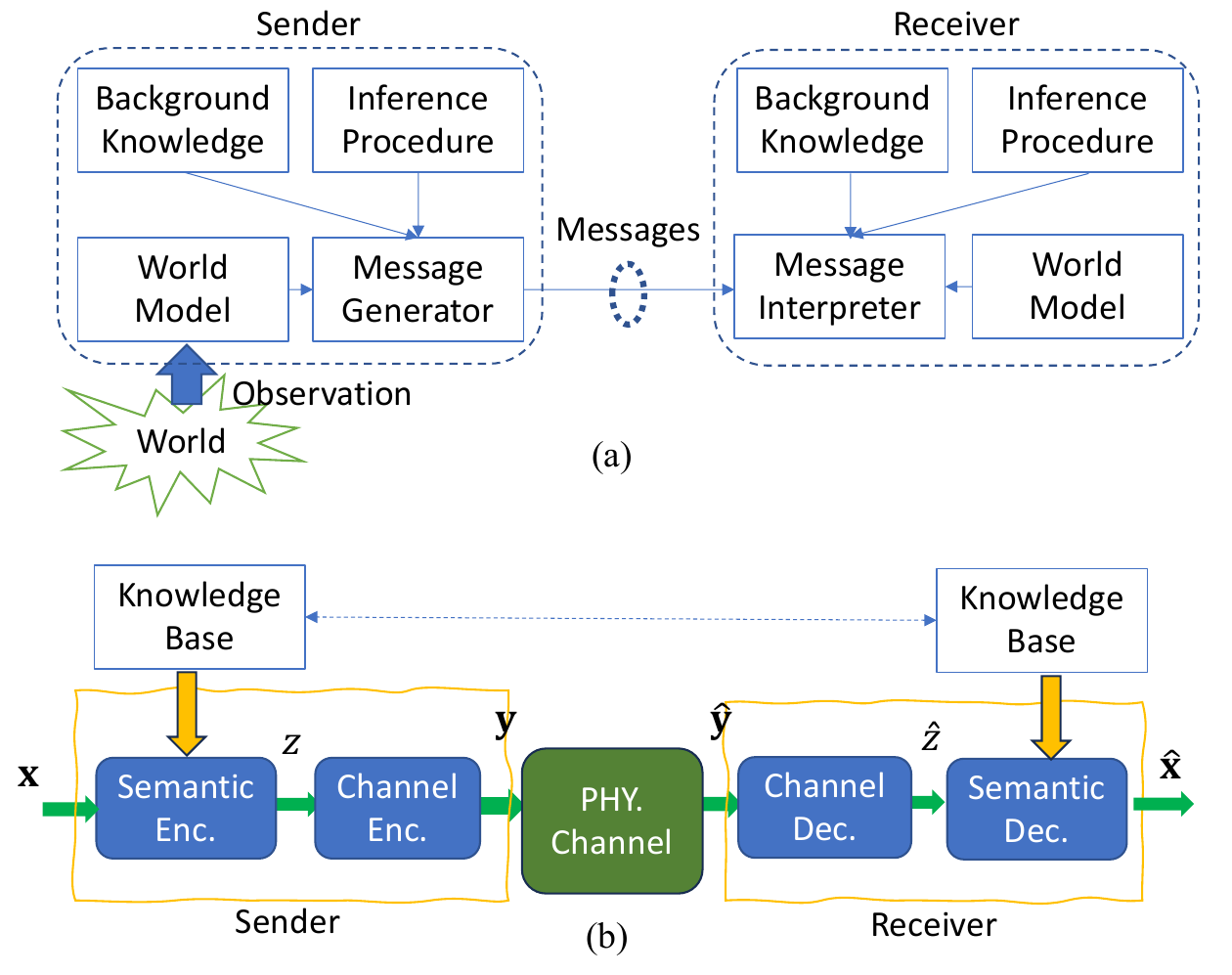} 
\end{center}
\caption{A model of SC system: (a) a generic model from \cite{Bao11}; (b) a model used for DeepJSCC 
(e.g., \cite{xie2021deep}).}
        \label{Fig:Fig1}
\end{figure}

The architecture for semantic communication proposed in \cite{Bao11}, which is illustrated in Fig.~\ref{Fig:Fig1} (a), is versatile and serves as a model for various approaches. For example, in \cite{xie2021deep} for DeepJSCC, background knowledge is utilized to extract given sentences, and the semantic encoder is conceptualized as a combination of the inference procedure and message generation within the context of text transmissions as shown in Fig.~\ref{Fig:Fig1} (b), where $\bx$ is a text input. 

The output of the semantic encoder, denoted by $\bz$, is referred to as the semantic representation. It is a discrete variable and an element of the semantic space, $\cZ$, i.e., $\bz \in \cZ$. In DeepJSCC, it is expected to have
\be
\log_2 |\cZ| < \sH (\bx) =
\uE[ - \log_2 (\bx)] \le \log_2 |\cX|
    \label{EQ:ZH}
\ee
so that the semantic representation can be used to go beyond the compression limit given by the entropy of the original signal in the context of source coding \cite{CoverBook}.

In \cite{nemati2023vq} and \cite{fu2023vector}, vector quantization is employed to quantize and encode the semantic feature vector, which is the output of the semantic encoder, to ensure reliable transmissions over noisy channels.

\subsection{Latent Space and Task-Oriented Communication}

According to the latent space theory \cite{Jiang2023}, each $\bx$ can be associated with an intention, denoted by $c$, where $c \in \cC$ is a discrete variable and $\cC$ is the set of all intentions. Thus, we have
\be 
p(\bx) = \sum_{c \in \cC} p(\bx\,|\, c) \pi (c),
\ee 
where $\pi (c)$ represents the prior distribution of $c$. Thanks to the sparsity property for well-trained models, for a given $\bx$, the associated intention can be found with a high probability, i.e.,
$$
\Pr(c\,|\, \bx) = \left\{ 
\begin{array}{ll}
1 - \epsilon, & \mbox{if $c = c_0$} \cr 
\epsilon, & \mbox{otherwise,} \cr 
\end{array}
\right.
$$
where $0 \le \epsilon \ll 1$ and $c_0$ represents the correct intention associated with $\bx$. As a result, when the semantic encoder is trained with a sufficiently large dataset, we expect that $\Pr(c\,|\, \bz)$ is also sufficiently high for $c = c_0$.

The latent space theory can be applied to task-oriented semantic communication \cite{Gunduz23}. For example, let $\cT (\bx)$ denote a task or test function with input $\bx$, and consider a classification task. Here, $\cT (\bx) \in \cC$ indicates that the output space of the task function consists of discrete variables, each associated with a label corresponding to the input $\bx$.
If the transmitter lacks sufficient computing resources to perform the task, it can send $\bx$ or its representation to the receiver with sufficient computing power.
In task-oriented semantic communication, $\bz$ can be transmitted so that $\hat \bx$ is generated at the receiver from the received semantic or latent variable, $\bz$. Then, it is expected to have
\be
\Pr(\cT(\bx) \neq \cT (\hat \bx) ) \geq 1 - \epsilon^\prime,
    \label{EQ:TT}
\ee
where $\epsilon^\prime \ll 1$ and 
$\Pr(\cT(\bx) \neq \cT (\hat \bx) )$ represents the probability that the outcomes of the task with the original signal and the generated signals at the receiver differ. This probability is not linked to the quality of the original signal $\bx$ in the task, but rather to the quality of the generated signal in terms of its proximity to the task's requirements. In other words, as long as $\bz$ can effectively convey semantic information for the given task or test without requiring $\bx \approx \hat{\bx}$, we can deem that semantic communication is successful when $\epsilon$ is sufficiently small. 

Of course, provided that the quality of the original signal is sufficiently good so that the error probability of the task, e.g., the classification error probability, is small, i.e.,
$$
\Pr(\cT(\bx) = c_0) \ge 1 - \delta,
$$
where 
$c_0$ represents the correct class label, and $\delta$ represents the acceptable classification error probability, which is sufficiently small. We expect that the receiver can also have a low classification error probability as follows:
$$
\Pr(\cT(\hat \bx) = c_0) \ge 1 - \delta - \epsilon^\prime.
$$

\subsection{Generative-Model based Semantic Communication}

In \cite{grassucci2024generative}, the use of generative models for semantic communication is investigated. This process involves reconstructing the signal at the receiver through a series of semantic conditioning steps with different semantic representations, potentially offering varying qualities of reconstruction. This flexibility allows for adaptation to different channel conditions, thereby enabling varying levels of reconstruction quality. As a result, compared to approaches based on DeepJSCC, it can be more flexible and adaptable to diverse communication scenarios, offering enhanced robustness and reliability in data transmission.

\section{Don'ts: Issues to Avoid}
\label{sec:donts}

\subsection{Direct Performance Comparisons with Conventional Communication}

Semantic communication aims to send the semantic representation that can be used to generate a signal at the receiver that might have the same meaning as the original signal, $\bx$. As a result, the generated signal, $\hat \bx$, is not necessarily similar to $\bx$ in terms of Euclidean distance or have a small MSE. In this sense, direct comparisons with conventional communication become meaningless. 

To comprehend this issue, let's analyze the first inequality in \eqref{EQ:ZH}. In the context of lossy compression, the entropy can decrease compared to that of the original signal, albeit with added distortion, which is often assessed through rate-distortion theory \cite{Berger71}. Consequently, achieving a transmission rate lower than the entropy of the original signal isn't unique to semantic communication; it can also be accomplished in lossy communication. However, the MSE, a widely-used distortion measure, cannot be directly applied to semantic communication. For example, as in \eqref{EQ:TT}, the classification error rate can be a performance measure in semantic communication.
Hence, direct comparisons between lossy compression and semantic communication (over a noiseless channel) cannot be made, as they necessitate different distortion measures.

\subsection{Semantic Communication for All}

Semantic communication might be seen as a special form of communication that can be useful in certain cases. For example, DeepJSCC can find semantic representations that can be expressed with a much smaller number of bits than $\sH(\bx)$ (i.e., beyond the limit of conventional source coding). However, since the semantic encoder and decoder are trained with a certain dataset that might be a set of samples drawn from a distribution, $p(\bx)$, if an input that is a sample drawn from a different distribution, the performance of semantic communication would be poor. In other words, semantic communication is not universal and can be used for specific cases. 

Another illustrative example highlighting the limitations of semantic communication emerges in task-oriented semantic communication scenarios. Let us consider a scenario where there are $K$ distinct tasks denoted by $\cT_k: \cX \to \cC$. As the number of tasks, $K$, increases, the challenge of finding a low-dimensional semantic representation $\bz$ capable of achieving consistently high success rates across diverse tasks becomes increasingly arduous.

In such cases, it is likely that the reconstructed signal $\hat \bx$ must closely approximate the original signal $\bx$ to ensure high success rates across all tasks. This tendency suggests that semantic communication may tend towards conventional communication methods, which typically focus on minimizing MSE. In essence, unless the number of distinct tasks is limited, the perceived advantage of task-oriented semantic communication in terms of reducing transmission rate may diminish.

\subsection{Impact on Physical Layer}

Machine learning techniques can be applied to the design of the physical layer, as discussed in \cite{OShea17}. Specifically, the autoencoder architecture can be utilized to design the transmitter and receiver for a given channel. This perspective can be extended to semantic communication through DeepJSCC, as demonstrated in \cite{xie2021deep}. Consequently, one may consider the potential impact of semantic communication on physical layer design and the development of new designs.

%it might be possible to investigate the potential impact of semantic communication on physical layer design.

In general, since semantic communication is based on semantic representations to convey the meaning of given signals or sources, it can be seen as extended source coding techniques with unconventional performance measures (e.g., learned perceptual image patch similarity (LPIPS) in \cite{Nam2023LanguageOrientedCW}). On the other hand, the physical layer may need to be used as a common platform that can support various applications with different kinds of source coding. Additionally, the physical layer is mainly responsible for managing physical channels, regardless of the types of sources and their encoding approaches. Essentially, to the physical layer, it is likely that semantic representations would be treated as another form of data packets to be reliably transmitted over noisy channels.

In other words, while it would be possible to consider a dedicated physical layer for semantic communication, the utility of such a dedicated physical layer would be limited. This does not, however, imply that there is a limited gain when the physical layer can be effectively designed to deliver semantic representations. By employing source-aware channel coding techniques, such as unequal error protection codes \cite{Masnick67} \cite{Hagenauer99}, it would be possible to design a dedicated physical layer (including channel coding) optimized for the characteristics of semantic representations.

%\subsection{Scalability??}
%The scalability ... arguable ...

\section{Dos: Issues to Investigate}
\label{sec:dos}

\subsection{Application of Rate Distortion Theory}

While the MSE is a typical measure in the rate distortion theory \cite{CoverBook} \cite{Berger71}, any distortion function can be used. Thus, it is possible to find the trade-off between the rate and distortion for task-oriented semantic communication. 

With multiple tasks in  the task-oriented semantic communication, we can define the distortion measure as 
\be 
d (K) = \max_{k \in \{1, \ldots, K\}} \Pr(\cT_k (\bx) \ne \cT_k (\hat \bx)) .
\ee 
With this distortion measure, the rate-distortion problem can be formulated as follows \cite{CoverBook}:
\begin{eqnarray}
    & \min_{p (\hat \bx|\bx) } \sI (\bx; \hat \bx) & \cr 
    & \mbox{subject to} \ d (K) \le 1 - \epsilon. &
\end{eqnarray}
Then, the semantic encoder and decoder can be trained to solve the above optimization problem.

For simplicity, we assume that the channel is noiseless. Then, we have the following Markov chain: $\bx \to \bz \to \hat \bx$. 
Using the data processing inequality \cite{CoverBook}, it can be shown that
\begin{align}
\sI (\bx; \hat \bx) & \le \sI (\bx; \bz)  \cr  
& = \sH(\bz) + \sH(\bz\,|\,\bx) \le \sH(\bz). 
\end{align}
Thus, instead of using $\sI (\bx; \hat \bx)$, we can use its upper-bound, $\sH(\bz)$, to see the rate-distortion relation with
$$
\{\mbox{Rate}, \ \mbox{Distortion} \} = \{\sH(\bz), d (K)\}.
$$
In task-oriented semantic communication, once the semantic encoder and decoder are trained with datasets, we can see that $\sH(\bz)$ is the number of bits to represent (or transmit) $\bz$ and $d(K)$ is the maximum error rate among $K$ different tasks.

%{\color{red}
%[Some Results ...]
%}

\subsection{Security}

In semantic communication, communicating parties need to train their semantic encoder and decoder with shared datasets. This process can be seen as a signaling game \cite{Choi22}.

\subsubsection{A Signaling Game} \label{SS:Lewis}

In the Lewis signaling game \cite{Lewis69}, there are two players, namely the sender and receiver. For convenience, the sender and receiver are called Alice and Bob, respectively.
There are the following three variables:
\begin{itemize}
\item \emph{Types}: $T \in \cT = \{t_k,\ k = 1,\ldots, N_T\}$, is a random variable that is observed by Alice.
\item \emph{Signals}: $S \in \cS = \{s_m,\ m = 1,\ldots, N_S\}$, is a signal that Alice sends to Bob.
\item \emph{Responses}: $R \in \cR = \{ r_n, \ n = 1,\ldots, N_R\}$, is a response that Bob chooses.
\end{itemize}
Here, $\cT$, $\cS$, and $\cR$ denote the sets of types, signals, and responses, respectively, having the same cardinally $K$, i.e., $N=N_T=N_S =N_R$.

% For simplicity, $N$ is assumed to be the same for $T$, $S$, and $R$
% as those of signals and responses. 

In the Lewis signaling game, Alice chooses a signal $S$ to send to Bob, depending on a given type $T$ that is randomly generated from a distribution $\pi$. In this game, Alice moves first, i.e., sending a signal, and then Bob moves next, i.e., receiving the signal and choosing a response, $R$. 
The payoff is given as:
\begin{align} 
u = \left\{
\begin{array}{ll}
1, & \mbox{if $R = T$;} \cr 
0, & \mbox{otherwise.} \cr 
\end{array} 
\right.
    \label{EQ:payoff}
\end{align}
Alice and Bob have mappings. The mapping at Alice is $S = \phi (T)$,
while that at Bob is $R = \theta (S)$. In order to maximize the payoff, Alice and Bob need to choose the mappings such that 
$T = R = \theta (\phi (T))$. There are $N!$ possible sets of mappings to maximize the payoff, including $s_n = \phi(t_k)$ and $r_k = \theta (s_k)$ for $k = 1,\ldots, K$. 

This signaling game is employed to model semantic communication in \cite{Choi22}. In the context of semantic communication, the types are the intentions and the signals are the semantic representations. The dataset to train the encoder and decoder is a set of realizations of types from a distribution $\pi(T)$, ${\mathbf T}_{\rm train} = \{t(1),
\ldots, t(L)\}$, where $L$ is the number of samples and $t(l) \in \cT$. This set of realizations, ${\mathbf T}_{\rm train}$, or the payoff in \eqref{EQ:payoff} has to be available to the receiver for training the decoder that is associated with the encoder.

\subsubsection{Secure Semantic Communication}

\begin{figure}[h]
\begin{center}
\includegraphics[width=\columnwidth]{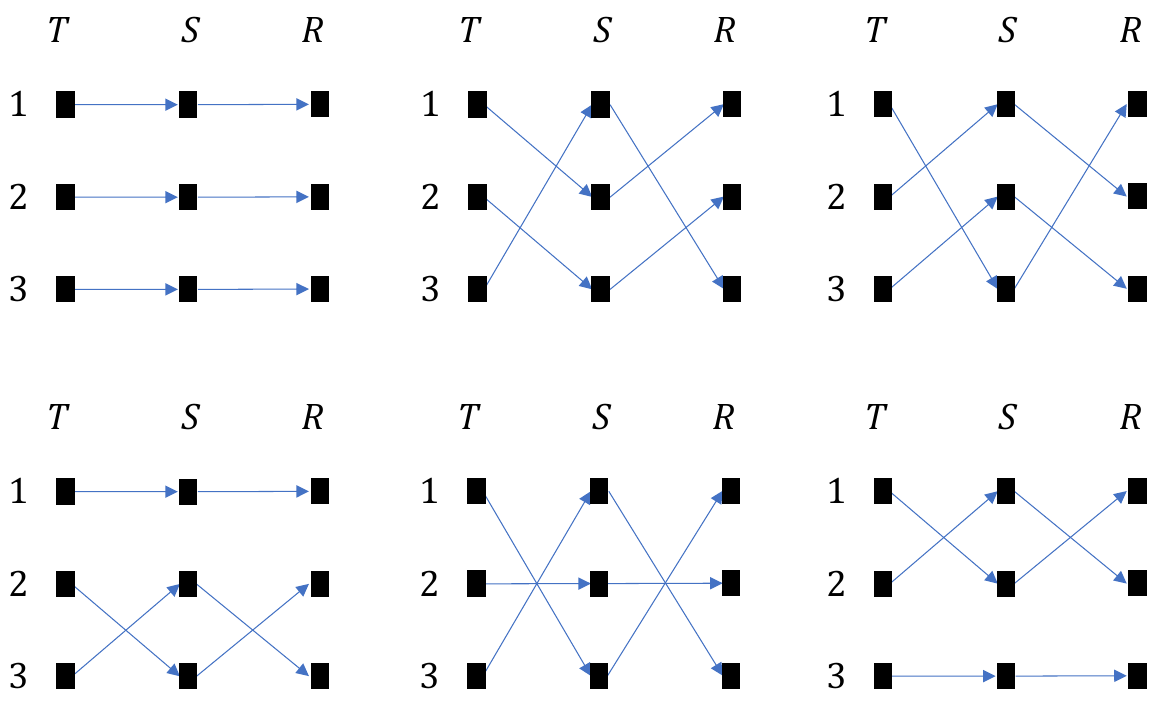} 
\end{center}
\caption{6 different mappings that maximize the payoff when $N = 3$ in the Lewis signaling game.}
        \label{Fig:Eqs3}
\end{figure}

A signaling game model provides a framework for secure semantic communication. Typically, the mapping rules for the encoder and decoder can converge to one of multiple equilibria, each maximizing the average payoff (e.g., as discussed in Subsection~\ref{SS:Lewis}, there are $N!$ equilibria in the Lewis signaling game), depending on the training dataset. This implies that there is no unique equilibrium, leading to a strong association between the encoder and decoder in each equilibrium state. Consequently, a pair of encoder and decoder trained differently lacks this association, resulting in the decoder's inability to accurately map the signal to its intended meaning. For example, when $N = 3$, there are $N! = 6$ possible mapping rules as illustrated in Fig.~\ref{Fig:Eqs3}. When the encoding and decoding rules are not associated, the receiver may not correctly decode the type. However, with differently associated encoder and decoder, the receiver can decode correctly for certain types in some cases. Additionally, since the encoding rule is fixed, it might be possible for an eavesdropper to understand the rule once a sufficient number of samples are given. This suggests that randomization and equivocation techniques might be employed to improve the security level.

\begin{figure}[h]
\begin{center}
\includegraphics[width=\columnwidth]{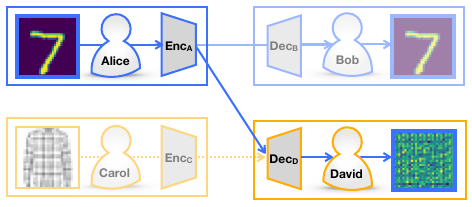} 
\end{center}
\caption{Two different pairs of communicating parties, Alice-Bob and Carol-David, trained under heterogeneous datasets (MNIST and Fashion-MNIST).}
        \label{Fig:Xpairs}
\end{figure}

Furthermore, if different datasets are used to train each pair of communicating parties, a receiver may not be able to decode the semantic representation transmitted by a sender from another pair. For instance, consider two pairs of communicating parties illustrated in Fig.~\ref{Fig:Xpairs}: one pair consists of Alice (sender) and Bob (receiver), and the other pair consists of Carol (sender) and David (receiver). The former pair is trained with MNIST, while the latter pair is trained with Fashion-MNIST. Consequently, when David eavesdrops on the signal transmitted to Bob by Alice, he fails to correctly generate a signal conveying the meaning of the original signal.

However, the study in \cite{choi2023semantics} suggests that partial re-training is feasible when different receivers need to comprehend the transmitted signals. To synchronize semantics across multiple neural transceivers, a distributed learning-based solution is proposed, which utilizes split learning (SL) and partial neural network (NN) fine-tuning techniques. 

Consequently, by training with different datasets or exploiting multiple equilibria, a certain level of security can be achieved, as discussed above. However, the security aspects of these approaches have not been thoroughly studied. Specifically, possible attack methods and their complexities  need to be investigated.

\subsection{Privacy via Generative Models}

In generative-model based semantic communication, there may be instances where certain features of the original signals are not precisely reproduced at the receiver. For example, while the poses and outlines of passengers can be reliably generated, variations in facial features may occur when transmitting vehicle dash camera images. This discrepancy highlights a notable advantage of generative-model based semantic communication in terms of privacy preservation. Unlike traditional methods that often require separate anonymization techniques to protect sensitive information, the inherent nature of generative models can offer a built-in privacy capability.

As discussed in \cite{grassucci2024generative}, the semantic conditioning procedure plays a crucial role in generative-model based semantic communication, as it enhances the quality of generated signals with each additional semantic conditioning step. Consequently, there may be a trade-off between signal quality, which entails providing more details, and introducing ambiguity for anonymization purposes. This implies that semantic conditioning can be optimized to maximize signal quality while maintaining a certain level of anonymization for privacy.

\subsection{Datasets and Training}

In semantic communication, the availability of datasets for training is crucial, as it forms the foundation of the communication process. This necessity introduces associated costs, both in terms of acquiring and maintaining datasets as resources. Furthermore, there is a significant computing cost incurred during the training phase. Consequently, the requirements for datasets and computing resources in semantic communication distinguish it from conventional communication methods, potentially offsetting the advantages it offers, such as significant reductions in bandwidth requirements.

However, it is worth noting that datasets can be loaded through background communication processes, particularly during periods of low network traffic. Moreover, training can be scheduled efficiently, and communication and computing resources can be managed effectively through distributed machine learning techniques. As a result, achieving cost-efficient dataset acquisition and training in terms of communication and computing becomes important.

%[resource management is usually for radio bandwidth, but now extended to datasets in semantic communication]

%\subsection{Training as Services}

\section{Conclusions}

In this paper, we discussed key aspects of semantic communication outlining its dos and don'ts. Semantic communication is not intended to replace conventional communication methods, which may still excel at minimizing bit-wise errors and ensuring universal interoperability.
Instead, it is worth exploring the unique capabilities and potential of semantic communication, such as its task-specific adaptation as well as security and privacy guarantees. Furthermore, it is crucial to address practical issues associated with implementing semantic communication within conventional communication system architectures, which involves identifying additional communication and computing costs as well as reusing and advancing existing physical layer designs.

\label{sec:conc}

\section*{Acknowledgement}
{\small{This research was supported by an IITP grant funded by the Korean government (MSIT) Information Technology Research Center (ITRC) support program (IITP-2023-RS-2023- 00259991).}}

\bibliographystyle{ieeetr}
\bibliography{ref}

\begin{thebibliography}{10}

\bibitem{Nam2023LanguageOrientedCW}
H.~Nam, J.~Park, J.~Choi, M.~Bennis, and S.-L. Kim, ``Language-oriented
  communication with semantic coding and knowledge distillation for
  text-to-image generation,'' in {\em IEEE Int. Conf. on Acoustics, Speech, and
  Signal Process. (ICASSP)}, 2023.

\bibitem{Grassucci2023GenerativeSC}
E.~Grassucci, S.~Barbarossa, and D.~Comminiello, ``Generative semantic
  communication: Diffusion models beyond bit recovery,'' {\em ArXiv preprint:
  arXiv:2306.04321}, 2023.

\bibitem{Grassucci2023DiffusionMF}
E.~Grassucci, C.~Marinoni, A.~Rodriguez, and D.~Comminiello, ``Diffusion models
  for audio semantic communication,'' in {\em IEEE Int. Conf. on Audio, Speech,
  and Signal Process. (ICASSP)}, 2024.

\bibitem{Wijesinghe2023DiffGODG}
A.~Wijesinghe, S.~Zhang, S.~Wanninayaka, W.~Wang, and Z.~Ding, ``{Diff-GO}:
  Diffusion goal-oriented communications to achieve ultra-high spectrum
  efficiency,'' {\em ArXiv preprint: arXiv:2312.02984}, 2023.

\bibitem{Wu2023CDDMCD}
T.~Wu, Z.~Chen, D.~He, L.~Qian, Y.~Xu, M.~Tao, and W.~Zhang, ``{CDDM}: Channel
  denoising diffusion models for wireless semantic communications,'' {\em ArXiv
  preprint: arXiv:2309.08895}, 2023.

\bibitem{grassucci2024generative}
E.~Grassucci, J.~Park, S.~Barbarossa, S.-L. Kim, J.~Choi, and D.~Comminiello,
  ``Generative ai meets semantic communication: Evolution and revolution of
  communication tasks,'' {\em ArXiv preprint: arXiv:2401.06803}, 2024.

\bibitem{Bao11}
J.~Bao, P.~Basu, M.~Dean, C.~Partridge, A.~Swami, W.~Leland, and J.~A. Hendler,
  ``Towards a theory of semantic communication,'' in {\em 2011 IEEE Network
  Science Workshop}, pp.~110--117, 2011.

\bibitem{xie2021deep}
H.~Xie, Z.~Qin, G.~Y. Li, and B.-H. Juang, ``Deep learning enabled semantic
  communication systems,'' {\em IEEE Transactions on Signal Processing},
  vol.~69, pp.~2663--2675, 2021.

\bibitem{CoverBook}
T.~M. Cover and J.~A. Thomas, {\em Elements of Information Theory}.
\newblock NJ: John Wiley, second~ed., 2006.

\bibitem{nemati2023vq}
M.~Nemati, J.~Park, and J.~Choi, ``{VQ-VAE} empowered wireless communication
  for joint source-channel coding and beyond,'' in {\em GLOBECOM 2023 - 2023
  IEEE Global Communications Conference}, pp.~1--6, 2023.

\bibitem{fu2023vector}
Q.~Fu, H.~Xie, Z.~Qin, G.~Slabaugh, and X.~Tao, ``Vector quantized semantic
  communication system,'' {\em IEEE Wireless Communications Letters}, vol.~12,
  no.~6, pp.~982--986, 2023.

\bibitem{Jiang2023}
H.~Jiang, ``A latent space theory for emergent abilities in large language
  models,'' 2023.

\bibitem{Gunduz23}
D.~Gündüz, Z.~Qin, I.~E. Aguerri, H.~S. Dhillon, Z.~Yang, A.~Yener, K.~K.
  Wong, and C.-B. Chae, ``Beyond transmitting bits: Context, semantics, and
  task-oriented communications,'' {\em IEEE Journal on Selected Areas in
  Communications}, vol.~41, no.~1, pp.~5--41, 2023.

\bibitem{Berger71}
T.~Berger, {\em Rate Distortion Theory: A Mathematical Basis for Data
  Compression}.
\newblock Prentice-Hall electrical engineering series, Prentice-Hall, 1971.

\bibitem{OShea17}
T.~O’Shea and J.~Hoydis, ``An introduction to deep learning for the physical
  layer,'' {\em IEEE Transactions on Cognitive Communications and Networking},
  vol.~3, no.~4, pp.~563--575, 2017.

\bibitem{Masnick67}
B.~Masnick and J.~Wolf, ``On linear unequal error protection codes,'' {\em IEEE
  Transactions on Information Theory}, vol.~13, no.~4, pp.~600--607, 1967.

\bibitem{Hagenauer99}
J.~Hagenauer and T.~Stockhammer, ``Channel coding and transmission aspects for
  wireless multimedia,'' {\em Proceedings of the IEEE}, vol.~87, no.~10,
  pp.~1764--1777, 1999.

\bibitem{Choi22}
J.~Choi and J.~Park, ``Semantic communication as a signaling game with
  correlated knowledge bases,'' in {\em 2022 IEEE 96th Vehicular Technology
  Conference (VTC2022-Fall)}, pp.~1--5, 2022.

\bibitem{Lewis69}
D.~K. Lewis, {\em Convention: A Philosophical Study}.
\newblock Cambridge, MA, USA: Wiley-Blackwell, 1969.

\bibitem{choi2023semantics}
J.~Choi, J.~Park, S.-W. Ko, J.~Choi, M.~Bennis, and S.-L. Kim, ``Semantics
  alignment via split learning for resilient multi-user semantic
  communication,'' 2023.

\end{thebibliography}

%\clearpage

\end{document}